\documentclass[amsmath,amssymb,twocolumn,showpacs]{revtex4}

\usepackage{graphicx}
\usepackage[usenames]{color}

\begin{document}

\title{Anisotropic charge dynamics in detwinned Ba(Fe$_{1-x}$Co$_x$)$_2$As$_2$}
\author{A. Dusza$^{1*}$, A. Lucarelli$^{1*}$, F. Pfuner$^{1}$, J.-H. Chu$^2$, I.R. Fisher$^2$ and L. Degiorgi$^1$} \affiliation{$^1$Laboratorium f\"ur
Festk\"orperphysik, ETH - Z\"urich, CH-8093 Z\"urich,
Switzerland}
\affiliation{$^2$Geballe Laboratory for Advanced Materials
and Department of Applied Physics, Stanford University, Stanford, California
94305-4045, USA and Stanford Institute for Materials and Energy Sciences, SLAC National Accelerator Laboratory, 2575 Sand Hill Road, Menlo Park, California 94025, U.S.A.}

\date{\today}

\begin{abstract}
We investigate the optical conductivity as a function of temperature with light polarized along the in-plane orthorhombic  $a$- and $b$-axes of Ba(Fe$_{1-x}$Co$_x$)$_2$As$_2$ for $x$=0 and 2.5$\%$ under uniaxial pressure. The charge dynamics at low frequencies on these detwinned, single domain compounds tracks the anisotropic $dc$ transport properties across their structural and magnetic phase transitions. Our findings allow us to estimate the dichroism, which extends to relatively high frequencies. These results are consistent with a scenario in which orbital order plays a significant role in the tetragonal-to-orthorhombic structural transition.

\end{abstract}



\maketitle

Understanding the interplay of structural, magnetic and superconducting phases is a major endeavor of ongoing research in the general field of complex interacting systems. The iron-pnictide superconductors \cite{kamihara,rotter} provide the most recent playground in which to address this challenging issue. The non-superconducting parent compounds undergo an antiferromagnetic transition into a broken-symmetry ground state at $T<T_{N}$, which is always preceded by or coincident with a tetragonal-to-orthorhombic structural distortion for temperatures $T\le T_{s}$ \cite{delacruz,lester}. This latter transition implies a two-fold electronic symmetry \cite{tanatar}, which for a range of dopings coexists with superconductivity and long range magnetic order \cite{lester,chu2}. In order to develop a comprehensive theoretical description of these materials, it is particularly important to understand the origin of the structural transition, and the physical properties of the material on the underdoped side of the phase diagram. However, the presence of dense structural domains which form below $T_s$ hides the intrinsic anisotropy of the orthorhombic phase in the FeAs ($ab$) plane \cite{tanatar}, such that naturally twinned samples present only an average of the intrinsic anisotropy, from which little detailed information can be extracted.  

It has recently been demonstrated that single crystals of Ba(Fe$_{1-x}$Co$_x$)$_2$As$_2$ and also the parent compounds BaFe$_2$As$_2$, SrFe$_2$As$_2$ and CaFe$_2$As$_2$ \cite{tanatar2,chudw} can be detwinned by application of uniaxial pressure, thus allowing investigations of the intrinsic in-plane anisotropy associated with the orthorhombic phase below $T_s$. Measurements of the resistivity of Ba(Fe$_{1-x}$Co$_x$)$_2$As$_2$ for currents flowing along the $a$ and $b$ axes ($\rho_a$ and $\rho_b$ respectively) reveal a surprisingly large anisotropy, reaching a maximum value $\rho_b/\rho_a\sim$ 2 for compositions close to the beginning of the superconducting dome \cite{chudw}. At higher temperatures, the applied uniaxial pressure naturally breaks the 4-fold symmetry of the tetragonal state, such that the observed resistivity anisotropy extends above $T_s$, revealing the presence of a substantial 'nematic susceptibility' \cite{nematic}. Consequently, the sharp phase transition at $T_s$ is changed into a broad crossover, due to the presence of the symmetry-breaking field \cite{chudw}.   

In this letter we focus our attention on two specific compositions, $x=0$ (i.e., the undoped parent compound BaFe$_2$As$_2$) and $x=0.025$ (i.e., an underdoped non-superconducting composition, close to the edge of the superconducting dome). The former ($x$=0) composition displays a small upturn in $\rho_b$ at $T_{s}$=$T_{N}$=135 K, while $\rho_a$ suddenly decreases at this temperature. The latter ($x$=0.025) composition is characterized by the onset of an insulating-like behavior of $\rho_b$ somewhat above $T_{s}$=98 K, for strained samples, while $\rho_a$ decreases with decreasing temperature in a metallic-like fashion \cite{chudw}. By means of the optical conductivity, obtained with light polarized along the $a$- and $b$-axes of samples under uniaxial pressure, we establish that the anisotropy of the charge dynamics extends to frequencies up to the mid- and near-infrared (MIR and NIR respectively) range, thus at energy scales well above $k_BT_{s}$. We discover that the effective metallic contribution to the excitation spectrum tracks the anisotropy observed in the temperature dependence of the $dc$ transport properties. Specifically, in the orthorhombic phase, the optical conductivity at far-infrared (FIR) frequencies along the $a$-axis is indeed more strongly enhanced with decreasing temperature relative to along the $b$-axis. In addition, a significant and progressive depletion of spectral weight is observed between 200 and 700 cm$^{-1}$ (previously ascribed to opening of a spin-density-wave gap in studies of twinned samples \cite{pfuner,lucarelli}) for $E\parallel b$ but \emph{not} for $E\parallel a$ on cooling below $T_N$. The most astonishing finding, however, consists in the substantial optical anisotropy that persists to even higher frequencies. We find two distinctive features, comprising a MIR band at 1500 cm$^{-1}$, which sits on the low frequency side of a pronounced NIR peak at 4300 cm$^{-1}$. Both of them are very anisotropic and undergo a distinctive reshuffling of spectral weight as a function of temperature. 

Similar to Ref. \onlinecite{chudw}, we developed a mechanical cantilever device that is able to de-twin crystals in situ, in this case coupled to our optical sample holder. The crystals, from the same growth batch used for the work of Ref. \onlinecite{chudw}, were cut and mounted on the device at room temperature such that in the orthorhombic phase the $a/b$ axes of the twinned samples would lie parallel to the direction in which the strain was to be applied. Uniaxial stress was applied by tightening a screw, drawing the cantilever down against the side of the crystal. Cooling samples for which uniaxial stress is applied in this manner results in a significantly larger population of domains for which the shorter $b$-axis is oriented along the direction of the applied stress, almost fully detwinning the crystals. The applied pressure is modest such that $T_N$ is unaffected, and can be adjusted over a limited range (up to approximately 5 MPa \cite{chudw}). Significantly, the cantilever detwinning device demonstrated in Ref. \onlinecite{chudw} leaves the (001) facet exposed, enabling us to perform optical reflectivity measurements on these detwinned, single domain samples. It was placed inside our cryostat, within the optical path of a Fourier interferometer, allowing optical measurements of the reflectivity $R(\omega)$ in the spectral range between 5 and 600 meV \cite{wooten,grunerbook}. Our data were complemented with room temperature measurements in the visible and ultra-violet spectral range, from 0.4 up to 6 eV. Light in all spectrometers was polarized along the $a$ and $b$ axes of the detwinned samples, thus giving access to the anisotropic optical functions. The real part $\sigma_1(\omega)$ of the optical conductivity was obtained via the well-established procedure of the Kramers-Kronig transformation of $R(\omega)$ by applying suitable extrapolations at low and high frequencies. For the $\omega\rightarrow 0$ extrapolation, we made use of the Hagen-Rubens (HR) formula ($R(\omega)=1-2\sqrt{\frac{\omega}{\sigma_{dc}}}$), inserting the $dc$ conductivity values ($\sigma_{dc}$) from Ref. \onlinecite{chudw}, while above the upper frequency limit $R(\omega) \sim \omega^{-s}$ (2$\le s \le$ 4) \cite{wooten, grunerbook}. 

Prior to performing optical experiments as a function of
the polarization of light, the electrodynamic response of the twinned (i.e., unstressed) samples was first checked 
with unpolarized light, consistently recovering the same
spectra previously presented in Ref. \onlinecite{lucarelli}. Although the cantilever device does
not permit  a precise control of the applied pressure, the uniaxial stress
was carefully increased enough to observe
optical anisotropy, which was verified to disappear when
the pressure was subsequently released. The two compositions investigated
displayed overall similar features in their optical response. To avoid
repetition, raw data are shown only for $x=0$. A comparison of the temperature
dependence of the dichroism for the two compositions is presented after the
initial analysis. 

\begin{figure}[!tb]
\center
\includegraphics[width=6.5cm]{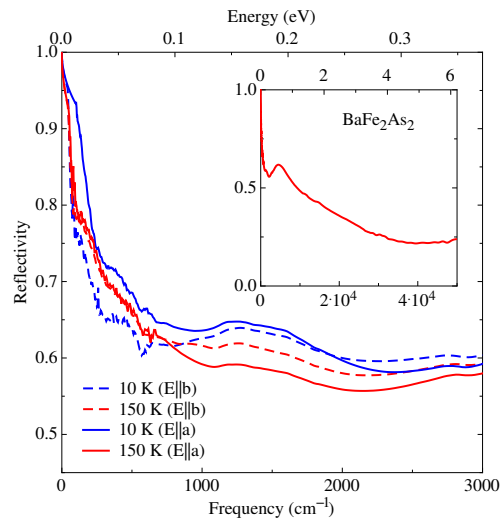}
\caption{(color online) Optical reflectivity $R(\omega)$ at 10 and 150 K of the parent compound in the FIR and MIR range for both polarization directions. Inset: $R(\omega)$ up to the ultra-violet spectral range.} \label{Refl.}
\end{figure}

Figure 1 displays $R(\omega)$ of the $x$=0 compound at $\omega<$3000 cm$^{-1}$ for light polarized along the $a$- and $b$-axes at two selected temperatures, well above and below the phase transitions. The $R(\omega)$ spectra tend to merge together in the energy interval between 5000 and 6000 cm$^{-1}$ so that neither a temperature nor a polarization dependence is observed above 6000 cm$^{-1}$. The inset illustrates the overdamped shape of $R(\omega)$ in the visible and ultraviolet spectral range, already recognized in the twinned specimens \cite{pfuner,lucarelli}. We first comment on the pronounced polarization dependence, leading to an optical anisotropy. The latter persists up to temperatures of at least 200 K, thus well above all phase transitions. In the absence of evidence for any additional phase transitions at higher temperatures, this behavior presumably reflects a substantial nematic susceptibility for $T>T_s$ \cite{nematic}. The optical anisotropy is even more enhanced as a function of temperature. At FIR frequencies and approaching the zero frequency limit, $R(\omega)$ increases with decreasing temperature along the $a$-axis, consistent with the metallic character of the $dc$ transport properties \cite{chudw}. On the contrary, along the $b$-axis, $R(\omega)$ displays a clear depletion at low temperatures in the energy range between 200 and 700 cm$^{-1}$, which is reminiscent of findings on the twinned samples \cite{pfuner,lucarelli}. $R(\omega\rightarrow0)$ is nevertheless consistent with the HR-expectation from the $dc$ transport. Another prominent feature in $R(\omega)$ is the strong polarization and temperature dependent band at about 1500 cm$^{-1}$. An attentive look to the data also reveals the interchange between the polarization dependence of $R(\omega)$ when crossing from high to low temperatures. Such an interchange occurs at $T_{s}$=$T_{N}$. 

The real part $\sigma_1(\omega)$ of the optical conductivity is shown in Fig. 2. Consistent with previous data \cite{pfuner,lucarelli}, $\sigma_1(\omega)$ is dominated by the strong absorption peaked at about 4500 cm$^{-1}$ and by the pronounced shoulder at 1500 cm$^{-1}$ on its MIR frequency tail. The inset of Fig. 2 shows $\sigma_1(\omega)$ in the FIR-MIR spectral range at 10 and 150 K, along both polarization directions. It clearly emphasizes the depletion of $\sigma_1(\omega)$ in the parent compound, occurring in FIR by lowering the temperature below $T_{N}$ and only affecting the $b$-axis response. Along the $a$-axis $\sigma_1(\omega)$ tends to acquire spectral weight and increases in FIR. This is consistent with the transport properties \cite{chudw}. For $x=0.025$ (not shown), we found a similar FIR behavior of the optical response, even though the transition at $T_N$ leads to less pronounced fingerprints, as already recognized in our previous work on twinned samples \cite{lucarelli}. Our findings thus demonstrate that the antiferromagnetic transition seems to partially gap the portion
of the Fermi surface pertinent to the $b$-axis response, while enhancing the metallic nature of the charge dynamics for the $a$-axis response.

\begin{figure}[!tb]
\center
\includegraphics[width=6.5cm]{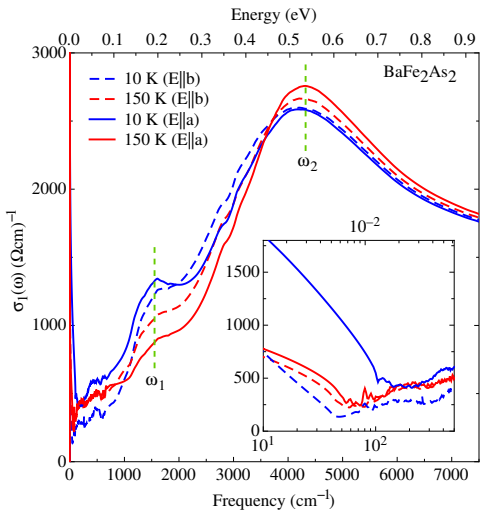}
\caption{(color online) Real part $\sigma_1(\omega)$ of the optical conductivity at 10 and 150 K of the parent compound in the MIR range for both polarization directions. The vertical dashed lines mark the frequencies $\omega_1$ and $\omega_2$ (see text). Inset: $\sigma_1(\omega)$ in the FIR-MIR range at 10 and 150 K for both polarization directions.} \label{sigma1}
\end{figure}

Another important result, evinced from our data, is the extension of the polarization as well as temperature dependence of the optical response up to energies, widely exceeding the energy scales set by the transition temperatures. The temperature dependence for each polarization direction is also well-reflected in the corresponding reshuffling of the spectral weight encountered in $\sigma_1(\omega)$ (Fig. 2). The FIR depletion of $\sigma_1(\omega)$ for $E\parallel b$ at $T<T_N$ removes spectral weight, which piles up in its incoherent part, principally in the MIR feature at 1500 cm$^{-1}$. For $E\parallel a$, the spectral weight essentially moves from energies around the peak at 4300 cm$^{-1}$ into the broad MIR shoulder at 1500 cm$^{-1}$ and also down to low energies into the effective metallic components (Fig. 1 and inset Fig. 2). A detailed analysis of the excitation spectrum for both polarization directions within the same Drude-Lorentz approach, introduced in our earlier study of the twinned compounds \cite{lucarelli}, is left to a forthcoming publication. Nonetheless, we anticipate here that the scattering rate of the itinerant charge carriers (i.e., the width of the effective (Drude-like) metallic response in $\sigma_1(\omega)$, inset Fig. 2) increases along the $a$-axis, while it decreases along the $b$-axis for $T<T_N$. This behavior is consistent with expectations for the given magnetic order, which is antiferromagnetic along the longer $a$-axis and
ferromagnetic along the shorter $b$-axis. Specifically, scattering from spin
fluctuations are anticipated to increase the scattering rate along the
$a$-axis. The results on the scattering rates of the itinerant charge carriers from
the phenomenological fit and the reshuffling of spectral
weight, pointed out above and occurring in the low energy intervals of $\sigma_1(\omega)$, uniquely determine the temperature dependence of the $dc$ transport properties \cite{chudw}. To account for the $dc$
resistivity anisotropy, it appears that anisotropy in the Fermi surface
parameters outweigh the anisotropy in the scattering rate that develops
below $T_N$.

\begin{figure}[!tb]
\center
\includegraphics[width=6.5cm]{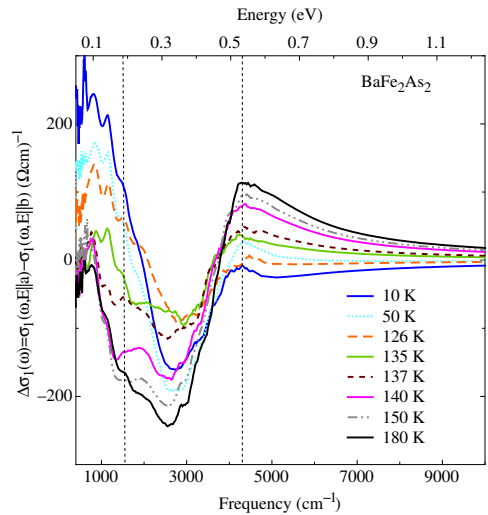}
\caption{(color online) Relative difference $\Delta\sigma_1(\omega)$ (dichroism) of the optical conductivity with polarization of light along the $a$- and $b$-axis at selected temperatures of the parent compound. The vertical dashed lines mark the frequencies $\omega_1$ and $\omega_2$ (see text and Fig. 2).} \label{dichroism}
\end{figure}

In order to emphasize the relevant polarization dependence, we calculate the difference $\Delta\sigma_1(\omega)=\sigma_1(\omega,E\parallel a)-\sigma_1(\omega,E\parallel b)$ at all measured temperatures. $\Delta\sigma_1(\omega)$, shown in Fig. 3, is somehow representing an estimation of the dichroism and turns out to be very prominent in the MIR and NIR ranges. At this point, it is worth establishing the compelling comparison with the anisotropy of the $dc$ transport properties, defined as $\frac{\Delta\rho}{\rho}$=$\frac{2(\rho_b-\rho_a)}{(\rho_b+\rho_a)}$ \cite{devereaux}. To this end, we first select two characteristic frequencies, identifying the position of the peaks in $\sigma_1(\omega)$; namely, at $\omega_1$=1500 (1320) cm$^{-1}$ and at $\omega_2$=4300 (5740) cm$^{-1}$ for $x=0$ (0.025) (Fig. 2). $\Delta\sigma_1(\omega)$ as a function of temperature at those two frequencies is reproduced in Fig. 4, together with $\frac{\Delta\rho}{\rho}$. We generalize here our overall arguments by including the case for $x$=2.5$\%$ as well. It is astonishing that the temperature dependence of the dichroism at selected energies tracks remarkably well the $dc$ anisotropy in both compounds. However, whereas the anisotropy in the $dc$ resistivity $\rho_b/\rho_a$ is larger for $x$= 0.025 than for $x$=0, the dichroism at selected frequencies (Figs. 3 and 4) is larger for the $x$=0 sample than for the sample with $x$=0.025. This doping-dependence needs to be studied in greater detail to make sure that it is not an artifact of the degree of detwinning of the particular samples used for this initial study. Even so, this result is certainly suggestive since the observed dichroism at high energies appears to have a similar doping-dependence as the lattice orthorhombicity (i.e., $a-b$/$a+b$) \cite{prozorov}. Our data
might thus establish a causal link between an electronic
effect (mainly involving high energy scales) and the structural transition. In contrast, $dc$ transport properties probe the low energy
quasiparticles, the behavior of which reflects details of the electronic
structure at the Fermi level.

\begin{figure}[!tb]
\center
\includegraphics[width=6.5cm]{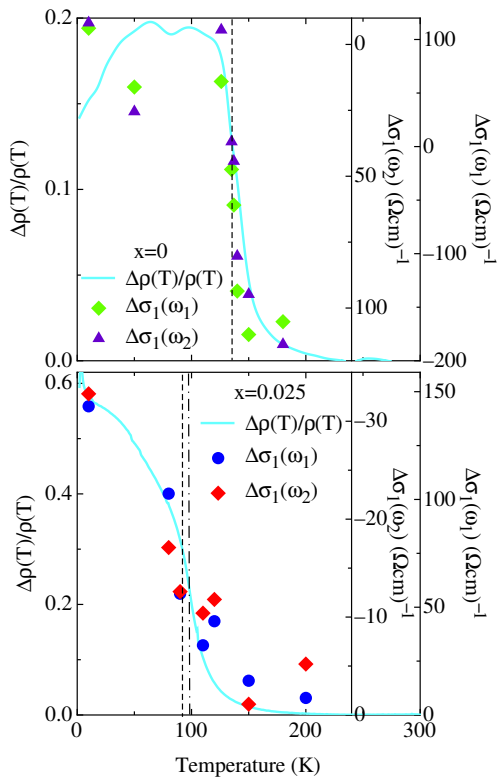}
\caption{(color online) Temperature dependence of the dichroism $\Delta\sigma_1(\omega)$ for $x$=0 and 0.025 at the selected frequencies $\omega_1$ and $\omega_2$ (Fig. 2) compared to the $dc$ resistivity anisotropy $\frac{\Delta\rho}{\rho}$ (see text) \cite{chudw}. The vertical dashed and dashed-dotted lines mark the magnetic and structural phase transitions at $T_{N}$ and $T_{s}$, respectively.} \label{anisotropy}
\end{figure}

In an attempt to shed light on the orthorhombic electronic anisotropy in the iron-pnictides, an approach based on orbital order was recently put forward \cite{kruger,chen,lee,lv,devereaux}. Orbital order can arise from an instability that promotes an electron in a local region from the $d_{xz}$ to the $d_{yz}$ orbital. Electron transfers leading to orbital polarization can locally break the tetragonal symmetry and couple to lattice distortions. Besides predictions of the magnetic order and of the resistivity anisotropy, consistent with experiment, the model based on orbital order also foresees an important dichroism \cite{devereaux}. While the X-ray absorption spectroscopy (XAS) could possibly reveal the orbital order parameter, our optical investigation can alternatively probe its consequence on the charge dynamics. The strongest polarization dependence of $\sigma_1(\omega)$ occurs at the two well-defined energy scales $\omega_1$ and $\omega_2$ (Fig. 2). They may represents optical transitions between states effectively characterized by an on-site energy splitting of about 0.3-0.4 eV, which is indeed compatible with the theoretical assumptions leading to the prediction of a linear dichroism in XAS \cite{devereaux}. Orbital order may also imply changes in the effective mass of the itinerant carriers, ultimately reflected in the spectral weight distribution (Fig. 2).

In summary, this initial survey of the optical response of detwinned single crystals of the representative iron-arsenide Ba(Fe$_{1-x}$Co$_x$)$_2$As$_2$ reveals a substantial in-plane anisotropy which extends to relatively high frequencies. The strong dichroism revealed by these data might be consistent with models based on orbital order which sets the stage for the structural as well as magnetic transitions. A dedicated theoretical calculation of the optical conductivity in our experimental configuration within that scenario is highly desired. Our optical findings may serve in order to test with great precision its implications when crossing over from $T>T_{s},T_{N}$ into the magnetic state.

The authors wish to thank S. Kivelson, T. Devereaux and D. Lu for fruitful discussions and J. Johannsen for valuable help in collecting part of the data. This work has been supported by the Swiss National Foundation for the Scientific Research
within the NCCR MaNEP pool. This work is also supported by the
Department of Energy, Office of Basic Energy Sciences under
contract DE-AC02-76SF00515.

$^{*}$ Both authors equally contributed to the present work.


\begin{thebibliography}{99}

\bibitem{kamihara} Y. Kamihara et al., \emph{J. Am. Chem. Soc.} \textbf{130}, 3296 (2008).

\bibitem{rotter} M. Rotter et al., \emph{Phys. Rev. Lett.} \textbf{101}, 107006 (2008).

\bibitem{delacruz} C. de la Cruz et al., \emph{Nature} \textbf{453}, 899 (2008).

\bibitem{lester} C. Lester et al., \emph{Phys. Rev. B} \textbf{79}, 144523 (2009).

\bibitem{tanatar} M.A. Tanatar et al., \emph{Phys. Rev. B} \textbf{79}, 180508 (2009).

\bibitem{chu2} J.-H. Chu et al., \emph{Phys. Rev. B} \textbf{79}, 014506 (2009) and references therein.

\bibitem{chudw} J.-H. Chu et al., \emph{cond-mat/1002.3364}.

\bibitem{tanatar2} M.A. Tanatar et al., \emph{Phys. Rev. B} \textbf{81}, 184508 (2010).

\bibitem{nematic} E. Fradkin et al., \emph{Annun. Rev. Condens. Matter Phys.} \textbf{1}, 7.1 (2010).

\bibitem{pfuner} F. Pfuner et al., \emph{Eur. Phys. J. B} \textbf{67}, 513 (2009).

\bibitem{lucarelli} A. Lucarelli et al., \emph{cond-mat/1004.3022}, and references therein.

\bibitem{wooten} F. Wooten, \emph{Optical Properties of Solids},
Academic Press, New York (1972).

\bibitem{grunerbook} M. Dressel and G. Gr\"uner, {\itshape Electrodynamics of Solids}, Cambridge University Press (2002).

\bibitem{devereaux} C.-C. Chen et al., \emph{cond-mat/1004.4611}.

\bibitem{prozorov} R. Prozorov et al., \emph{Phys. Rev. B} \textbf{80}, 174517 (2009).

\bibitem{kruger} F. Kr\"uger et al., \emph{Phys. Rev. B} \textbf{79}, 054504 (2009).

\bibitem{chen} C.-C. Chen et al., \emph{Phys. Rev. B} \textbf{80}, 180418(R) (2009).

\bibitem{lee} C.-C. Lee et al., \emph{Phys. Rev. Lett.} \textbf{103}, 267001 (2009).

\bibitem{lv} W. Lv et al., \emph{cond-mat/1002.3165}.


\end{thebibliography}
\end{document}